\documentclass[11pt,twoside]{article}
\usepackage{asp2006}
\usepackage{epsf}
\usepackage{graphics}
\usepackage{lscape}
\def\edcomment#1{\iffalse\marginpar{\raggedright\sl#1\/}\else\relax\fi}
\reversemarginpar

\begin{document}

\markboth{Hudson \& Li}{Flare/CME Properties}
\title{Flare and CME Properties and Rates at Sunspot Minimum}
\author{H.~S. Hudson and Yan Li}
\affil{SSL, UC Berkeley}

\begin{abstract}
The corona at solar minimum generally differs greatly from that during active times.
We discuss the current Cycle 23/24 minimum from the point of view of the occurrence of flares and CMEs (Coronal Mass Ejections).
By comparison with the previous minimum, the flare/CME ratio diminished by almost an order of magnitude.
This suggests that the environmental effect in flare/CME association differed in the sense that the Cycle 23/24  minimum corona was relatively easy to disrupt.
\end{abstract}

\vspace{-0.5cm}
\section{Introduction}

Periods of low activity reveal a less complicated coronal structure.
In addition far fewer flares and coronal mass ejections (CMEs) occur; when they do they produce effects that develop in simpler ways because of their simpler environments.
The Cycle 23/24 minimum is particularly interesting in this respect because of its extent, unprecedented during modern times.
This period has also seen the deployment of remarkable new observational capability, including the first stereoscopic observations of the corona.
In this paper we describe the present knowlege of occurrence patterns, both temporal and spatial, and attempt to discern whether or not this altered environment has resulted in systematic changes in the event properties, and if so what these changes imply in the physical mechanisms involved.

In discussing flare/CME occurrence patterns, one makes explicit or implicit assumptions about the nature of the referent.
What is a CME, and what is a flare?
The chosen answer to one of these questions obviously has a major influence on how one describes the
occurrence.
In this paper we will make the standard working hypothesis that the flare represents the luminous output of the process, whereas the CME \citep[e.g.,][]{2000JGR...10518169S} represents certain ejecta.
We adopt the GOES soft X-ray event to define ``flare,'' but note that this results in a strongly biased sample.
Ejecta come in two orientations: a plasma flow parallel to the field could be called a ``jet,'' while
a plasma flow perpendicular to the field results in the appearance of a set of expanding loops and could
readily be identified as at least a part of a CME.
These categories are not mutually exclusive, but in a true jet \citep[e.g.,][]{1996PASJ...48..123S} there is no appearance of the formation of new open magnetic flux.

This paper begins with a general discussion of occurrence distributions -- temporal, spatial, and joint flare/CME occurrence -- and then attempts to derive lessons from the recent remarkable solar minimum separating Cycles 23 and 24.

\section{Occurrence Patterns}

\subsection{Magnitude distributions}

Major flares and CMEs occur less frequently than minor ones.
This holds both for solar \citep{1956PASJ....8..173A,1993SoPh..143..275C} and stellar \citep[e.g.,][]{1989SoPh..121..375S} flares.
Flare occurrence distributions follow a flat power law \citep{1991SoPh..133..357H}, and in this they resemble many other natural phenomena.
The Gutenberg-Richter law for earthquakes is a well-known example; see Press (1978) 
\nocite{1978ComAp...7..103P}
for an interesting commentary on this phenomenon and its relationship to ``1/f noise,'' for example. 
Flares occurrence  probably follows a power-law distribution in energy release, but normally we deal with proxies such as soft X-rays because the complex, broad-band nature of flare energy release makes it hard
to characterize observationally.
Usually we identify this energy with a specific observable, such as the GOES class, but this invariably imposes a bias due to selection effects.
The GOES energy flux represents only about  one percent of the bolometric luminosity of a flare \citep[][Quesnel et al., these Proceedings]{2005JGRA..11011103E}.

CMEs differ from flares physically, in that the coronal magnetic field blows open and links a new part of the solar atmosphere to the solar wind temporarily (e.g., Gold, 1962).
\nocite{1962SSRv....1..100G}
This absorbs a major fraction of the energy release, both in the CME motions directly and also by the stretching the coronal magnetic field into a more energetic state, with the creation of a current sheet that may last for many days (e.g. Bemporad 2009).
\nocite{2009ApJ...701..298B}
A flare usually happens at the launch of a CME, but not always (Webb et al. 1998; Robbrecht et al. 2009).
\nocite{1998GeoRL..25.2469W,2009ApJ...701..283R}
The likelihood of a CME increases rapidly with the GOES class of the flare (Yashiro et al. 2005), 
\nocite{2005JGRA..11012S05Y}
such that no ``CME-less'' flares with GOES classes above X2 have yet been reported \citep{2007ApJ...665.1428W}.
This selection effect would imply that the CME distribution function of occurrence might have a different power law (or a different kind of best-fitting function), and presumably one that is still flatter than the flare occurrence distribution.
The flare distribution fails to converge when extrapolated above the observed energy range (Hudson 1991), and it is so flat that most of the luminous energy of flaring in the aggregate comes from the few most powerful events.

\begin{figure}
\plottwo{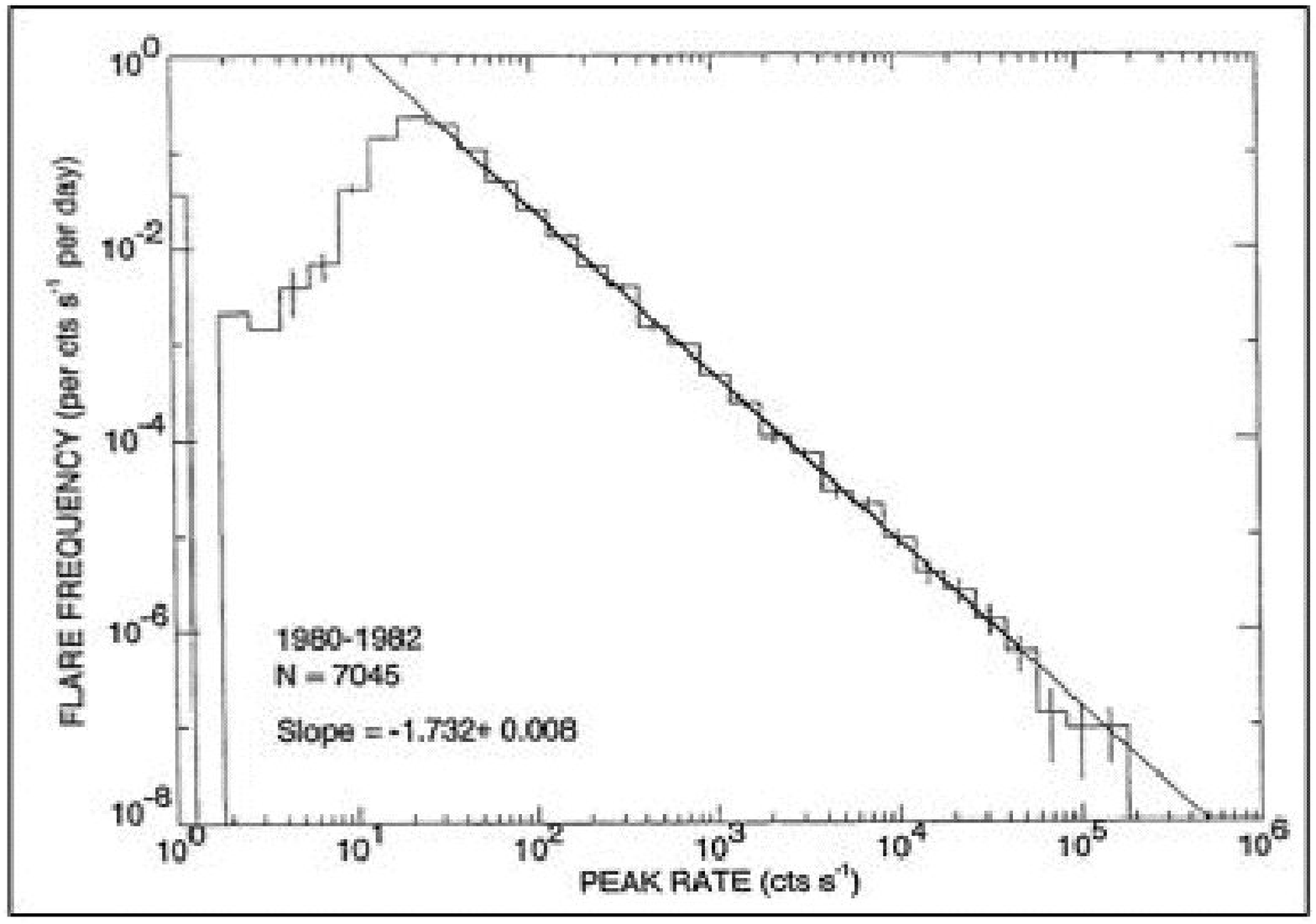}{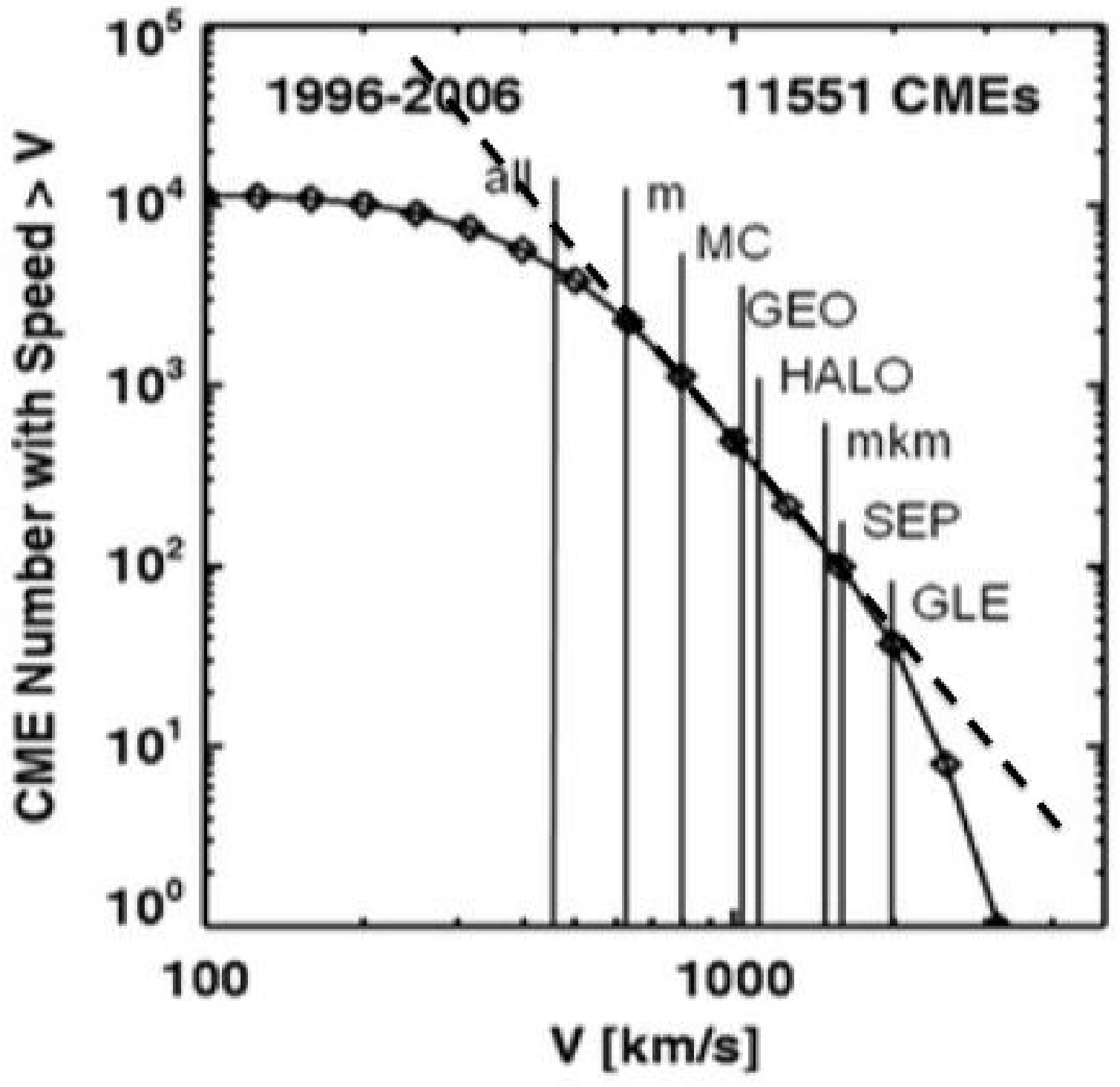}
\caption{\label{fig:crosby}
\textit{Left:} the \cite{1993SoPh..143..275C} differential distribution of flare magnitudes, for which the slope = 1.732~$\pm$~0.008; \textit{right:} integral distribution of CME velocities, with vertical lines denoting the average speeds of various classes of events \cite[for definitions, see][]{2009EM&P..104..295G}.
The dashed line has a slope of~3.6.
}
\end{figure}

Flare energy distributions (the number of flares per unit total energy, for example) can readily be 
estimated from the statistics of proxies such as the X-ray emission.
The distribution (Figure~\ref{fig:crosby} left) is a flat power law \citep{1991SoPh..133..357H,1993SoPh..143..275C}, such that the most powerful events dominate the total energy even considering extrapolation
to the microflare range \citep{1991SoPh..133..357H}.
The slope ($\alpha$ defined by the fit dN/dE~$\sim$~E$^{-\alpha}$) has the well-defined value 1.732~$\pm$~0.008 in the \cite{1993SoPh..143..275C} analysis.
 
The CME energy content consists of the kinetic energy of the moving mass, its gravitational potential energy, its enthalpy change, and its magnetic energy (e.g. Vourlidas et al. 2000).
\nocite{2000ApJ...534..456V}
The magnetic term is dominant but hard to assess, and so the proxy understanding of the total energy is not so clearly defined.
Figure~\ref{fig:crosby} (right) shows instead the integral distribution of CME velocities as obtained from the SOHO/LASCO CME catalog \citep{2009EM&P..104..295G}. 
This is interestingly different from the distribution of flare energies, in that it is much steeper.
The dashed line in Figure~\ref{fig:crosby} (right) has a slope of 3.6 for the integral distribution in $v$, 
and the CME energy might scale as $v^2$.
This strikingly different distribution law contrasts with that found for solar energetic particle fluences by \cite{1975SoPh...41..189V}, for which the slope is 1.15~$\pm$~0.05.
This is much flatter than even the flare distribution, and totally inconsistent with the CME distribution.
This points to interesting unexplored physics in the relationship between CMEs and the solar energetic particles.

The flatness of the power-law distribution of flare occurrence requires the existence of an upper cutoff
\citep[e.g.,][]{2007ApJ...663L..45H}
but this is hard to observe simply because events near the cutoff are so infrequent.
The long, homogeneous data sets from the Solar Maximum Mission/HXRBS  \citep{1993SoPh..143..275C} and Ulysses/GRB \citep{2009SoPh..258..141T} do not show obvious breaks in the power law, suggesting that the cutoff is no lower than about X10.
By studying the occurrence statistics of individual active regions, \cite{1997ApJ...475..338K} have shown that the power-law slope stays roughly the same but that the upper cutoff in fact varies.

\subsection{Statistics}

\bigskip\noindent
In a series of papers, M. Wheatland has described flare occurrence in terms of the waiting-time distribution,
ie the series of times between events \citep[e.g.,][]{2002SoPh..211..255W}.
The waiting-time distribution is an exponential function if the occurrence follows a simple Poisson distribution,
as expected if flares occur as independent events. 
If the flare occurrence rate varies, the waiting-time distribution develops a power-law tail.
To describe a long-term variation, one must allow the mean rate to vary, but Wheatland found that a piecewise constant rate would do, with the rate constant over time intervals given for example by a ``Bayesian block'' \citep{1998ApJ...504..405S} technique.

For the less frequent energetic events, the block length becomes longer, and at our present state of knowledge the extreme flares above the X10 level, where the downward break in the distribution function might occur, would be at a rate of a few events per cycle -- consistent with a Poisson distribution over this long time interval.

\subsection{Flare/CME relationships}

\begin{figure}
\plottwo{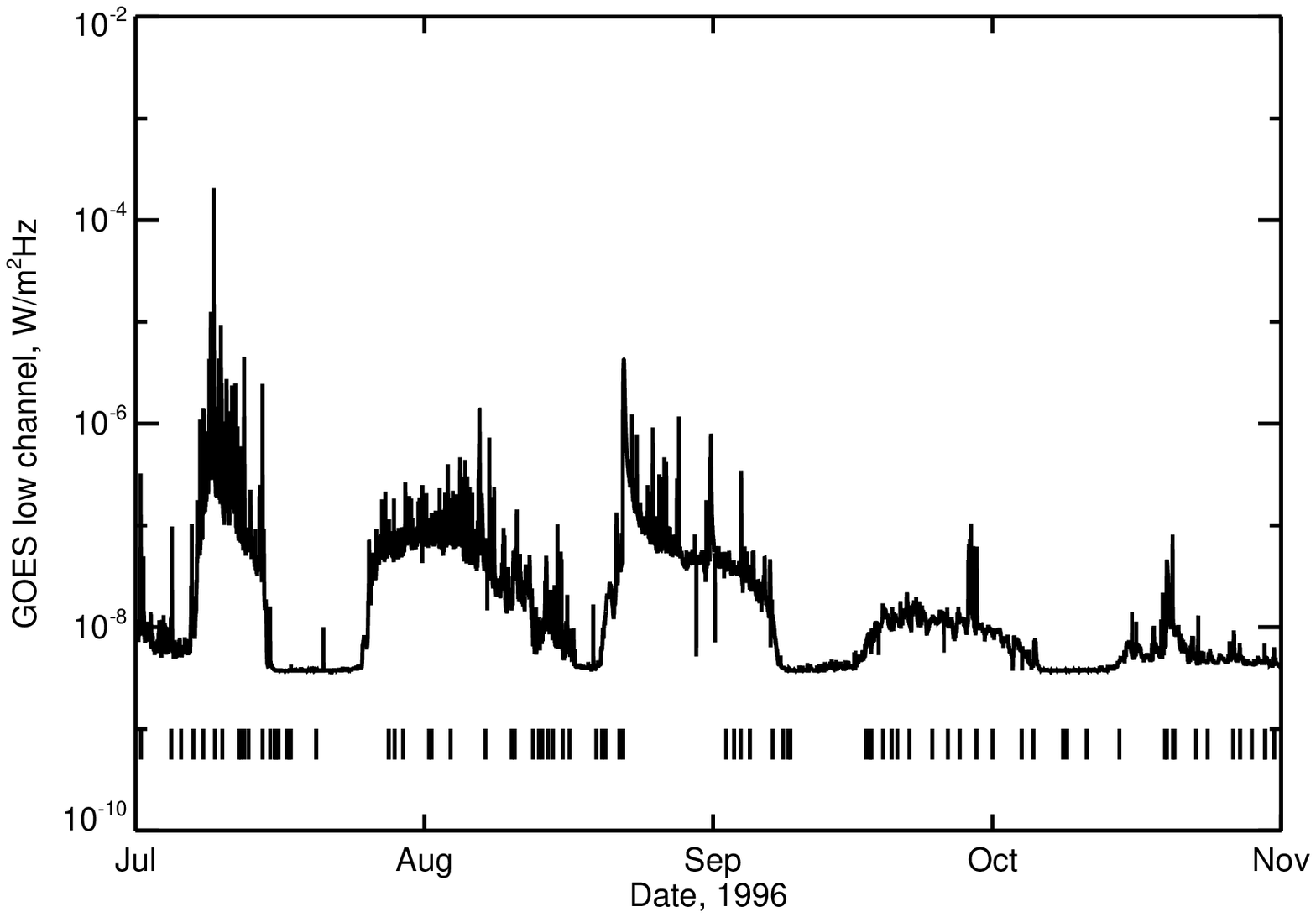}{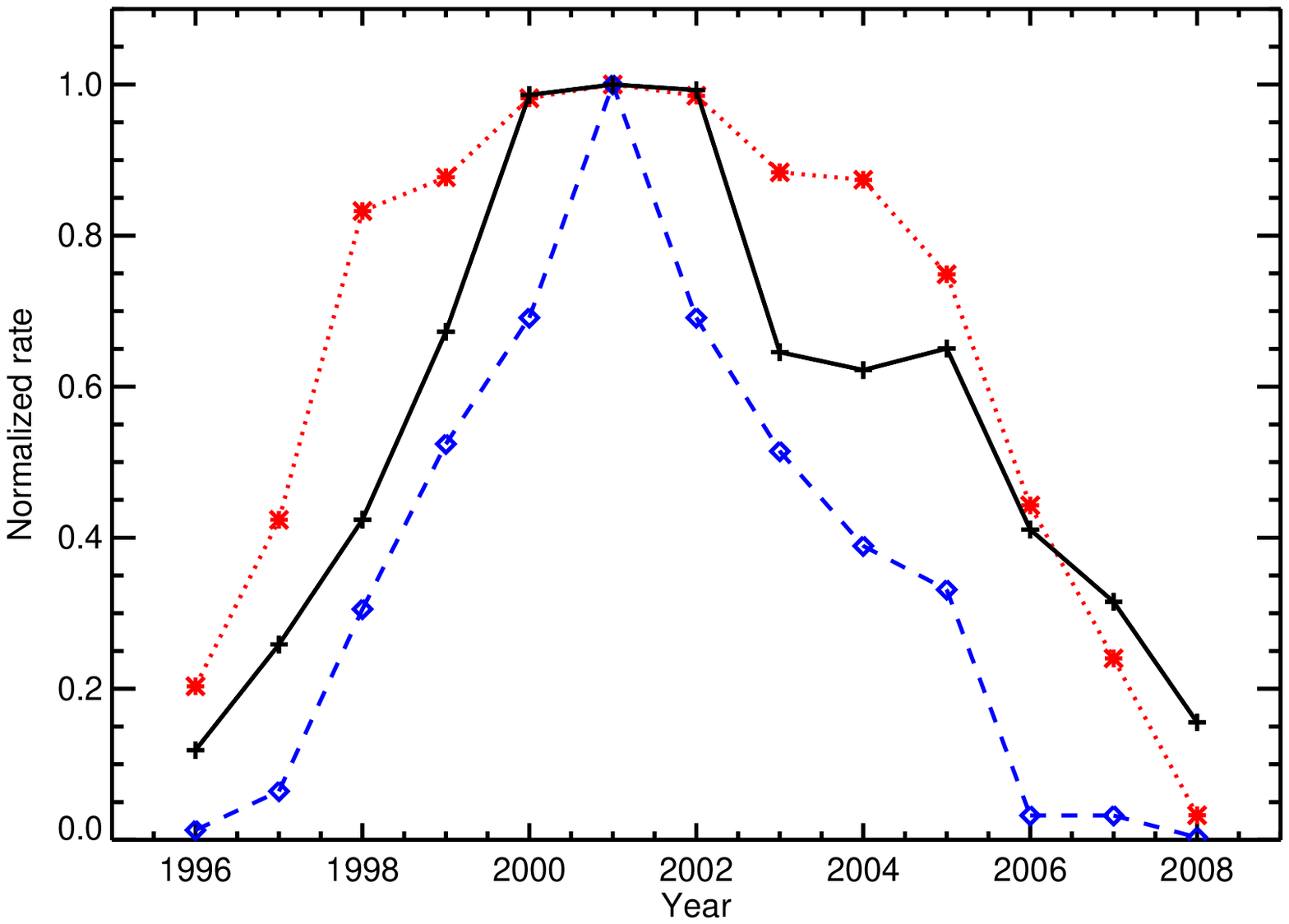}
\caption{\label{fig:cycle_plot}
\textit{Left:} GOES flux for the five rotations of the ``last best region'' of Cycle 22, beginning with its initial eruption. The tick marks at bottom show CME occurrence times.
\textit{Right:} Cycle 23, showing yearly averages of CMEs (black solid line) with widths $>$25$^\circ$,  and of GOES events (all events, red dotted line; M-class events only, blue dashed line).
All curves are normalized to the peak year (2001 in each case).
}
\end{figure}

Figure~\ref{fig:cycle_plot} (left) shows the joint CME and flare occurrence (via the GOES flux) for the ``last best region'' (NOAA 7983) of Cycle 22 \citep{1998ASSL..229..237H}.
This clearly shows the complexity of the correlation between flares and CME over five solar rotations.
One can see that CMEs continue even after flaring from the region has dramatically diminished: for each CME there would likely have been a flare, especially for the major flares in the first two rotations of this region, but they rapidly became less important following the initial X-class flare of 1996 July~6.
The CME pattern tends to show the front/back modulation so obvious in the soft X-rays, and also a tendency
for the limb passages to be enhanced presumably because of greater CME visibility then.

\begin{figure}
\plottwo{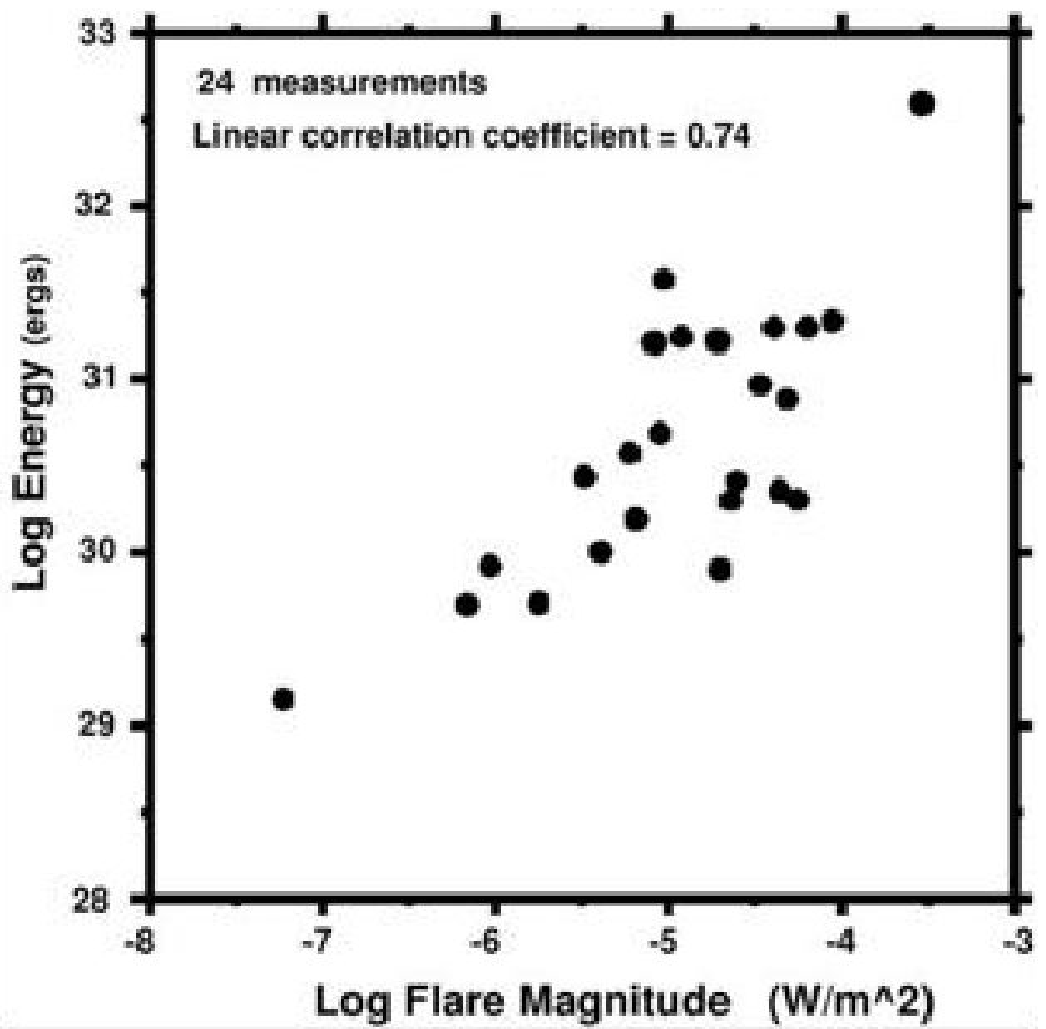}{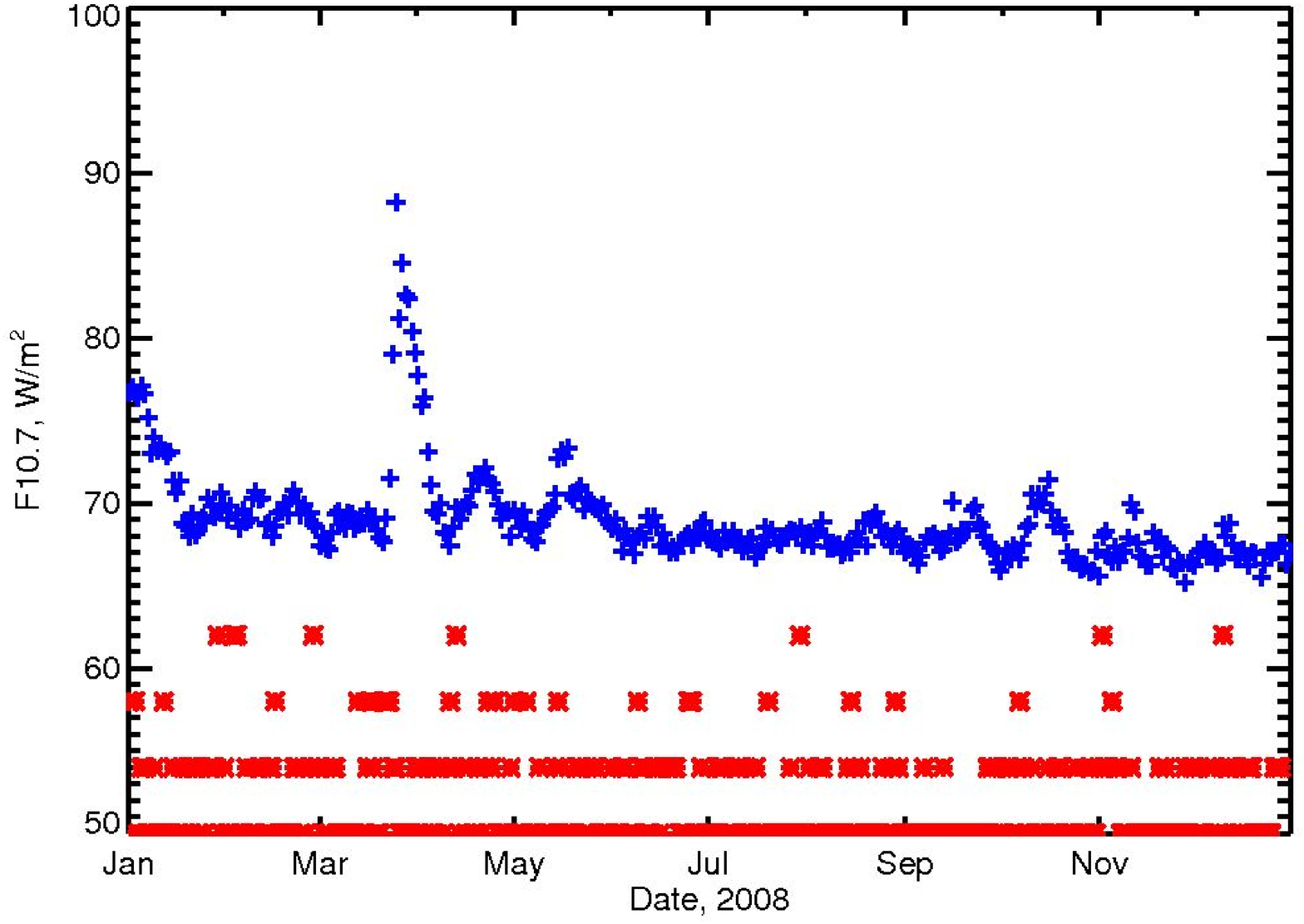}
\caption{\label{fig:corr}
\textit{Left:} association between CME kinetic energy and GOES magnitude, as derived for a set of limb flares by \cite{2004JGRA..10903103B}.
\textit{Right:} Occurrence of CMEs (red asterisks, in the range 0-4 events per day) during the Cycle 23/24 minimum year of 2008. 
The blue crosses show F10.7.
}
\end{figure}

In spite of the apparently loose association between flare and CME rates illustrated in Figure~\ref{fig:cycle_plot} (left), a physical relationship appears to exist for the most energetic events.
The evidence for this lies in the energetics, and in the timing.
The correlation in Figure~\ref{fig:corr} (left) shows the results of flare/CME energy comparison by \cite{2004JGRA..10903103B}.
These suggest a general correlation between the kinetic energy of the CME and the radiated energy of its associated flare.
Note that \cite{2005JGRA..11011103E} find comparable energies for the radiant energy and CME kinetic energy in two major events.

\cite{2005JGRA..11012S05Y} found that (20\%, 49\%, 91\%) of (C, M, X)-class flares had associated CMEs listed in the SOHO/LASCO CME catalog.
The remaining X-class flares include some that definitely do not have any CME association, and \cite{2007ApJ...665.1428W} infer from the circumstances of these flares that the coronal environment played the decisive role.
Apparently the flare/CME process has an energy threshold above which the corona can be forced open, or literally blown apart.
This points clearly to the limitation of CME occurrence by the influence of the overlying coronal field or solar wind.

How closely tied are the flare and CME aspects of the overall process?
The timing of flare and CME signatures often has been argued to imply a causal relationship one way or the
other.
The standard model ``explains'' flare loops in terms of the reconnection of field opened up in the initial part of the event.
This explanation makes little sense for events without eruptions, so it cannot be general and really should not be considered a standard model for the event as a whole.
When a CME does occur, though, it often begins with early signs -- filament activations many minutes before the flare impulsive phase were recognized many years ago, for example.
The acceleration phase of the CME appears to match the timing of the impulsive phase quite well
\citep{2001ApJ...559..452Z,2004ApJ...604..420Z}, and since this is where energy appears most intensely in mass motions and in both particle acceleration and flare luminosity, this provides strong evidence for a physical association.
This timing association had already been discovered indirectly via X-ray dimming \citep[e.g.,][]{1996AIPC..382...88H,1999ApJ...520L.139Z}, on the hypothesis that the sudden X-ray dimming results from the depletion
of the coronal volume needed to supply the CME mass.

Early studies of flare/CME relationships have had to suffer from geometrical problems.
For example, flares that occur behind the limb can still be associated with CMEs, so that a blind survey might find that almost half of the CMEs seemed to be ``flareless.''
This effect was detected as soon as statistics became good enough \citep{1979SoPh...61..201M} but often forgotten later on.
Another important effect is the angular dependence of the Thomson-scattering cross-section, which favors events
at the limb.
This in the past was often taken to justify the ``plane-of-the-sky'' approximation; nowadays one can almost always locate the disturbance in the lower atmosphere and thus pinpoint the location of the base of the CME.
Coronagraphs can detect CMEs originating from any point on the disk (or behind it), but with complicated responses depending on the geometry \citep{2006ApJ...642.1216V}.
The STEREO observations have succeeded to the extent that the directions of CME motions can be
measured to of order 10$^\circ$ \citep{2009SoPh..259..143X,2009SoPh..256..183T}.

\section{The Cycle 23/24 Minimum Period}

\subsection{Flare/CME event rates}

In this section we discuss the appearance of the Cycle 23/24 minimum period in terms of event occurrence.
The cycle minimum, as judged by the F10.7 index, appears to have been in late 2008.
Accordingly the SOHO time series (beginning in 1996) nicely embraces one complete maximum and two minima.
Figure~\ref{fig:cycle_plot} (right) shows a comparison of GOES flares with CME numbers for events tabulated in the LASCO catalog \citep{2009EM&P..104..295G} as having widths~$>$25$^\circ$ and speeds~$>$100~km/s.
We distinguish all GOES events from energetic ones (M~class; blue dotted line); the latter have less obscuration from major long-decay events and thus provide a more correct sample.
The red dotted line shows the total GOES event listings, and it indeed has a broader width in time as one would expect if weak flares were being under-counted at maximum.

At solar mimimum times the GOES under-counting effect for weak flares becomes less significant.
For the two minima shown (1996 and 2008) we find the GOES/CME count ratio to be 3.80~$\pm$~0.35 and
0.46~$\pm$~0.06, respectively, a highly significant difference.
We conclude that although flares basically ceased in 2008 (only one M-class and 8 C-class flares), CMEs continued to occur.
While most of the most energetic CMEs have flare associations, some spectacular ones come from the eruptions of filaments in quiet-Sun regions \citep[e.g.,][]{2006SSRv..123...13H}.
These events are not flares in the classical sense, although there are pale imitations (large, slowly-developing ribbon structures; vast soft X-ray arcades with low peak temperatures) of ordinary flare events \citep[e.g.,][]{1986stp..conf..198H}.
Accordingly we can attribute the base level of CME occurrence seen in Figure~\ref{fig:corr} at the cycle minima to events of this type.
Without this contribution the M-flare rate by itself, for example, correlates strongly with the CME occurrence frequency (Figure~\ref{fig:cycle_plot}, right), with allowance for a base rate of quiet-Sun CME formation.

The CME totals in the minimum years (1996, 2008) were (145, 190) respectively, roughly the same at about~14\% of the count in the intervening maximum year (2002); the amplitude of the solar cycle in annualized flare rate is much larger.
The Cycle 23/24 minimum count was marginally \textit{greater} than that of the Cycle 22/23 minimum, but 2008 was offset from the apparent minimum (December, according to the F10.7 index), and so the numbers are roughly comparable.
Figure~\ref{fig:corr} (right) shows that the CMEs that did occur in 2008 (again, selected from the LASCO/SOHO CME catalog for width $>$~25$^\circ$ and speed  $>$~100~km/s) did not obviously reflect the quiet-Sun variations apparent in F10.7.

\section{Conclusions}

The Cycle 23/24 minimum lasted longer than expected and exhibited unprecedented properties, at least as regards the modern era (see e.g. Svalgaard~\& Hudson in these Proceedings). 
The LASCO/SOHO CME catalog contains about the same number of events in 2008 the previous minimum year of 1996, even though major flares virtually ceased.
The observation that CMEs continue into solar minimum whereas flares do not, or else become less important, suggests the existence of a CME mechanism basically independent of the activity level.

The ratio of flare and CME occurrences was almost an order of magnitude (0.12~$\pm$~0.02) smaller than in the previous minimum, consistent with the concept that the weaker coronal magnetic field at minimum times is easier to disrupt. 
This environmental effect is so strong that some major CMEs can occur with minimal low-coronal visibility, i.e. no flare at all \citep[e.g.,][]{1998GeoRL..25.2469W,2009ApJ...701..283R}.

CME occurrence during the Cycle 23/24 minimum year of 2008 does not show any particular correlation with the small fluctuations in solar activity during that time, as judged by F10.7.
We expect that most of the 190 CMEs at our criterion during that time had a `flare' association in the most general sense, but that in many cases the actual event may have been hard to identify.
These events would be worth detailed study nevertheless in the sense that they represent simple disruptions of the coronal magnetic field from an elementary state, and thus may guide us to the basic physics of the eruption process better than a major flare-associated event during solar maximum.

\bigskip\noindent
{\bf Acknowledgements:}
This work was supported by NASA under grant NAG5-12878 (HSH), and by grants NNG06GE51G and NNX08AJ04G (YL).

\bibliographystyle{cspm-bib}
\bibliography{hudson-li} 

\end{document}